\PassOptionsToClass{layout=twocolumn}{achemso}
\documentclass{achemso}

\usepackage[version=3]{mhchem} 
\usepackage[english]{babel}
\usepackage[colorlinks=true, allcolors=blue]{hyperref}
\usepackage{amssymb}
\usepackage{relsize}
\graphicspath{{images/}}
\usepackage{multicol}
\usepackage{xr-hyper}

\SectionNumbersOn
\newcommand{\colorit}[1]{\textcolor{blue}{#1}}
\newcommand{\nohCGmodelName}{\text{AMARO}}

\makeatletter
\newcommand*{\addFileDependency}[1]{\typeout{(#1)}
\@addtofilelist{#1}
\IfFileExists{#1}{}{\typeout{No file #1.}}
}\makeatother
\newcommand*{\myexternaldocument}[1]{%
\externaldocument{#1}%
\addFileDependency{#1.tex}%
\addFileDependency{#1.aux}%
}
\myexternaldocument{SI} 

\author{Antonio Mirarchi}
\affiliation{Computational Science Laboratory, Universitat Pompeu Fabra,
Barcelona Biomedical Research Park (PRBB), Carrer Dr. Aiguader 88,
Barcelona, 08003, Spain.}
\author{Raúl P. Peláez}
\affiliation{Computational Science Laboratory, Universitat Pompeu Fabra,
Barcelona Biomedical Research Park (PRBB), Carrer Dr. Aiguader 88,
Barcelona, 08003, Spain.}
\author{Guillem Simeon}
\affiliation{Computational Science Laboratory, Universitat Pompeu Fabra,
Barcelona Biomedical Research Park (PRBB), Carrer Dr. Aiguader 88,
Barcelona, 08003, Spain.}
\author{Gianni De Fabritiis}
\affiliation{Computational Science Laboratory, Universitat Pompeu Fabra,
Barcelona Biomedical Research Park (PRBB), Carrer Dr. Aiguader 88,
Barcelona, 08003, Spain.}
\altaffiliation{Institucìo Catalana de Recerca i Estudis Avançats (ICREA), Passeig
Lluis Companys 23, Barcelona, 08010, Spain.}
\alsoaffiliation{Acellera Labs, Doctor Trueta 183, Barcelona, 08005, Spain.}
\email{g.defabritiis@gmail.com}

\title
  {AMARO: All Heavy-Atom Transferable Neural Network Potentials of Protein Thermodynamics}
\begin{document}
\begin{abstract}
All-atom molecular simulations offer detailed insights into macromolecular phenomena, but their substantial computational cost hinders the exploration of complex biological processes. We introduce Advanced Machine-learning Atomic Representation Omni-force-field
(\nohCGmodelName{}), a new neural network potential (NNP) that combines an O(3)-equivariant message-passing neural network architecture, TensorNet, with a coarse-graining map that excludes hydrogen atoms. \nohCGmodelName{} demonstrates the feasibility of training coarser NNP, without prior energy terms, to run stable protein dynamics with scalability and generalization capabilities. 
\end{abstract}

\section{Introduction}
Molecular events at the individual macromolecule level reveal emergent and collective macroscopic behaviors, emphasizing the need for a comprehensive and hierarchical approach to deeply investigate the complexities of biophysical complexes critical to all cellular functions. \cite{williamson2008cooperativity}. Over the past decades, the integration of advanced computational methods, with the latest hardware and software, and the increasing availability of experimental molecular structure data, has deeply transformed molecular biology and drug discovery  \cite{mccammon1977dynamics, hollingsworth2018molecular}. These enhancements, supported by significant advancements in theoretical frameworks, have evolved molecular simulations from simple proofs of concept to detailed \textit{in silico} studies of protein folding \cite{baxa2014loss, noe2009constructing} and dynamics \cite{muller2019characterization, mcgeagh2011protein}. 

\begin{figure*}
    \centering
    \includegraphics[width=\textwidth]{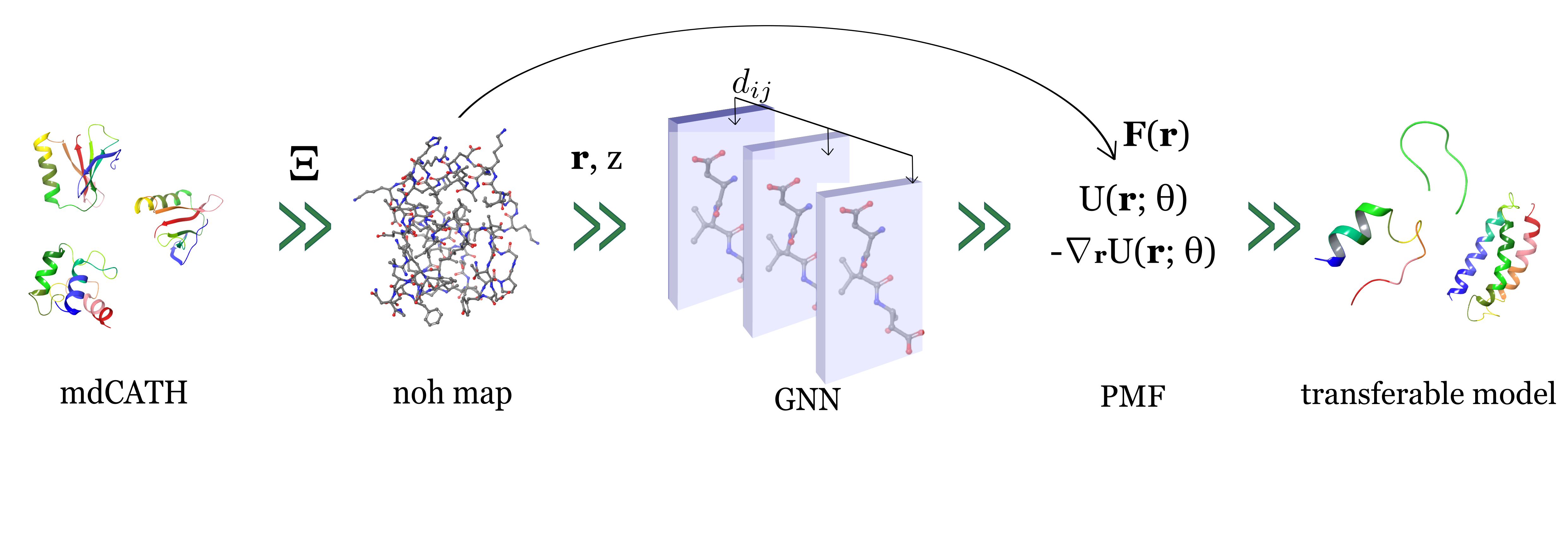}
    \caption{The pipeline for developing all-heavy atom NNPs is reported here. A CG map is applied to the mdCATH dataset \cite{mirarchi2024mdcath}, and an embedding $z$ for each domain is created. TensorNet\cite{simeon2023tensornet} is then trained using the CG data. Generalization and scale-up properties are evaluated on a set of four fast-folding proteins and larger domains in the final stage.} 
    \label{fig:pipeline}
\end{figure*}

Molecular dynamics (MD) simulations, in particular, serve as a powerful tool for capturing the behaviors of proteins and biomolecules at the atomic level, providing fine temporal resolution. However, classical all-atom MD simulations present multiple limitations due to the substantial increase in computational resources required for larger systems and extended time scales; furthermore, post-process and data analysis demands considerable effort, particularly in terms of human expertise and time \cite{chodera2014markov, plattner2017complete}. To investigate larger systems over an extended time scale, a leading approach involves reducing computational demands via coarse-grained (CG) simulations, where molecular systems are simulated using fewer degrees of freedom than those associated with the atomic positions \cite{Pak2018, noid2013perspective}. 

Several CG models have been developed, each tailored to optimize molecular simulations and capture critical biophysical features. The MARTINI model \cite{marrink2007martini, monticelli2008martini}, for instance,   excels due to its adaptability across a range of biomolecular systems, including membrane structure formation and protein interactions. Models such as AWSEM \cite{davtyan2012awsem} and UNRES \cite{liwo2019general} have been successful in simulating intramolecular protein dynamics, though they occasionally struggle to capture alternative metastable states. Similarly, Primo \cite{gopal2010primo, kar2013primo} and Rosetta \cite{das2008macromolecular} focus on specific molecular interactions and the design of protein structures and complexes, enhancing their accuracy in targeted applications.
Currently, there is no universally accepted theory that precisely determines the most effective coarse-graining mapping for any given system, which in general is related to the intended application and computational constraints \cite{foley2020exploring, boninsegna2018data}. Once a coarse-graining mapping has been established, different strategies exist to define the model energy function, either to reproduce the reference fine-grained statistic (bottom-up) \cite{jin2022bottom} or to match experimental observables (top-down) \cite{navarro2023top}. 
Neural network potentials (NNPs) demonstrate remarkable efficiency in rapidly learning accurate potential energy \cite{noe2020machine} and effectively model many-body atomic interactions \cite{behler2007generalized, wang2021multi}. These features are extremely beneficial for developing coarse-grained (CG) force-fields that need multi-body functional forms to accurately represent protein thermodynamics and incorporate implicit solvation effects.
Many CG-NNP models \cite{wang2019machine, husic2020coarse, chen2021machine, durumeric2023machine, charron2023navigating} have been presented in the past years, but most rely on prior energies for stability and accuracy. In the context of CG-NNPs, the prior energy terms are defined as contributions to the final energy prediction of a model that are predefined (either constant or via some function of the atomic labels) or based on physical principles, independent of the machine-learning model. Here we present the first version of our Advanced Machine-learning Atomic Representation Omni-force-field (\nohCGmodelName{}), a bottom-up CG-NNP without the need of prior terms to achieve stability and transferability. \\

\section{Material and methods}
A CG model typically comprises two main components: selecting the CG resolution (or mapping) and designing an effective energy function once the mapping has been established. Mapping schemes often draw from physical or chemical intuition. CG  sites may represent functional groups, residues, or monomers, or be tailored to a specific resolution. In this study, we adopt a no-hydrogen coarse-graining map, and a machine learning model is employed to learn the potential energy function \cite{kmiecik2016coarse}. Traditional force fields have historically treated hydrogens primarily as charge carriers due to their negligible Lennard-Jones interactions with heavier atoms. However, significant advances in implicit atom models, including the united atoms representation and implicit water machine learning potentials, illustrate the feasibility of accounting for these interactions using a many-body approach \cite{yang2006new,chen2021machine}.

\subsection{Neural Network Model} 
In bottom-up CG approaches the interactions between CG beads are determined based on a more detailed model, and the many-body potential of mean force (PMF) is used to describe the free energy landscape of a system as a function of a collective coordinate or reaction coordinate, which is in other terms a configurational free energy in a reduced space. The term "many-body" in the context of a CG model refers to the fact that the potential energy landscape is considered in terms of interactions between groups of atoms rather than individual atoms, and network models offer a straightforward approximation in this context \cite{noid2013perspective}. Our machine learning model of choice in this work is TensorNet\cite{simeon2023tensornet}, a new neural network architecture that integrates O(3)-equivariance in message-passing and utilizes rank-2 Cartesian tensor representations. O(3)-equivariant NNPs \cite{batzner20223, batatia2022mace, musaelian2023learning} ensuring that tensor outputs transform correctly under rotations and reflections. In practice, in this work, we only predict the scalar energy, therefore invariance would have been enough. TensorNet, using a Cartesian representation, does not add a significant extra cost in incorporating these features, while it might provide higher expressive power and accuracy in energy and force prediction \cite{duval2023hitchhiker}.

\subsection{No hydrogen CG map}
The selection of an optimal mapping approach for transitioning from a fine- to coarse-grained representation is a critical aspect of a CG model definition. An effective CG map should significantly reduce the computational burden of the all-atom model while preserving sufficient information to prevent an excessively flat energy surface. In this study, a no-hydrogen (\textit{noh}) and no-water coarse-grained map has been chosen to reduce the degrees of freedom by almost half compared to the all-atom counterpart and to align with the basic force aggregation method \cite{krämer2023statistically}. 

Consider a dataset $\mathcal{D}$ consisting of coordinate-force pairs obtained using an all-atom MD force field. The dataset has $M$ systems, each with a potentially different number of atoms. For each system, the coordinates are denoted by $\mathbf{r} \in \mathbb{R}^{3N}$ and the forces by $\mathbf{f} \in \mathbb{R}^{3N}$, where $N$ is the number of atoms in that particular system. A linear map operator $\Xi: \mathbb{R}^{3N} \rightarrow \mathbb{R}^{3n}$ is defined to map an all-atom conformation $\mathbf{r}$ to a coarse-grained conformation $\mathbf{R}$, where $n$ denotes the number of non-hydrogen atoms. This mapping is applied separately to each system, accommodating their atomic compositions. A paired coarse-grained force $\mathbf{F}$ is then considered for any conformation $\mathbf{R}$, and the force of the $i$-th \textit{noh-bead} $\mathbf{F}_i$ is defined as:

\begin{equation}
\mathbf{F}_i = \mathbf{f}_{\textit{ih}} + \sum_{j \in \mathbb{B}} \mathbf{f}_j,
\end{equation}

\noindent where $\mathbf{f}_{\textit{ih}}$ represents the force of the $i$-th heavy atom, and $\mathbb{B}$ denotes the set of hydrogen atoms bonded to the $i$-th heavy atom in the fine-grained representation. Accounting for the noh coarse-graining map, this approach uniquely embeds each noh-bead by considering both its corresponding heavy atom and the number of bonded hydrogen atoms. A final set of 12 embedding values, as outlined in Table  \colorit{S1}, was obtained enhancing the model's ability to differentiate between various electronic hybridizations. This selection empowers the model to more effectively learn the atoms' properties and molecular geometry. 

\subsection{Neural network training} 
As demonstrated in prior research, the acquisition of the many-body PMF involves minimizing the mean-squared deviation between a CG candidate force field and atomistic forces appropriately mapped. This method, known as variational force matching \cite{izvekov2005multiscale}, establishes a robust approach to effectively learning the PMF, laying the foundation for the exploration of intricate biomolecular interactions\cite{kmiecik2016coarse,noid2008multiscale}. At CG resolution the force matching method becomes more complicated since the training data contain less information than their atomistic counterparts: energies are not available, and forces are noisy \cite{durumeric2023machine}. To enhance the quality of input data, we employed the \textit{basic force aggregation} method in dataset preparation. Coupled with the noh-CG map, this approach 
significantly improves PMF learning \cite{krämer2023statistically}. In essence, the model now learns the force of a \textit{noh-bead} as the sum of the forces acting on its heavy atoms and the constrained (i.e., 'bonded') hydrogen atoms. Notably, this model introduces a groundbreaking feature in the CG-NNP field moving from delta-learning to directly learning the forces acting on particles, no prior terms are considered. So, the energy function is parametrized by the network parameters $\theta$ which are optimized to minimize the mean-squared deviation between predicted and labeled forces via the loss function

\begin{equation}
\resizebox{\linewidth}{!}{$
L_{FM}(\theta) = \frac{1}{\sum_{k=1}^K N_k} \sum_{k=1}^{K} \sum_{i=1}^{N_k} \left\| -\nabla_{\mathbf{R}_i^k} \widetilde{U}(\mathbf{R}^k; \theta) - \mathbf{F}_i^k \right\|^2 ,$}
\label{loss_function}
\end{equation}

\noindent where $N_k$ is the number of beads in conformation $k$, and $K$ is the total number of conformations in a batch. Predicted forces are obtained as the negative gradient of the potential energy $\widetilde{U}$ with respect to the \textit{noh}-bead coordinates $\mathbf{R}$, and $\mathbf{F}$ represents the labeled coarse-grained forces.

\subsection{Dataset}\label{sec: dataset}
The mdCATH dataset \cite{mirarchi2024mdcath} was the basis for applying the coarse-grained mapping approach. Specifically, the initial data were processed to retain only the heavy atoms' coordinates and forces, with a basic force aggregation map applied to the latter. Additionally, the \textit{z} dataset within each HDF5 file was modified to serve as the embedding for each system. A series of filters were implemented to exclude certain domains based on the following criteria: 1) domains containing more than 150 residues, 2) domains comprising more than 1000 noh-atoms, or 3) domains with less than 50\% combined helix and sheet fractions. Various temperatures (320~K, 348~K, 379~K, 413~K, 450~K) were considered to facilitate the model's learning of atomic proximity or separation dynamics. Since we have trained the model using multiple temperatures, we are not expected to reproduce the energetics at 350 K exactly. Nonetheless, we find the agreement reasonable. The reason for training at multiple temperatures is that it increases the variability of the training data.
In total, 2,834 domains and more than 26 million conformations were selected from the mdCATH dataset after considering the filters and temperatures as described above.

The final dataset used to train the model was obtained by applying a stride of 25 to the 26M residual conformations, resulting in an approximate split of 900,000 conformations for training, 50,000 for validation, and 100,000 for testing.

The model’s scalability was then tested on larger mdCATH domains, specifically those with more than 150 residues and a combined helix and sheet content greater than 50\%.

At the same time, transferability was assessed using four fast-folding proteins: Chignolin, Trp-cage, Villin, and $\alpha$3D; sequence similarity details are reported in Table \colorit{S2}.

\subsection{AMARO Molecular Simulations}\label{MolSims}
All MD simulations reported in this work, involving \nohCGmodelName{}, were conducted with the mass of each CG-bead calculated as the combined mass of the heavy atoms and hydrogen atoms that constitute the bead.

\subsection{Markov State Models}\label{MSM_Section}
In this study, we analyze the dynamics of CG simulations using Markov State Models (MSMs) \cite{noe2008transition, husic2018markov, pan2008building} available in HTMD \cite{doerr2016htmd} and compare them with those from corresponding all-atom simulations. MSMs partition the entire dynamics of a system into \textit{n} discrete states and are particularly suited for systems that exhibit Markovian behavior—where future states depend only on the current state without memory of the past. We construct a transition probability matrix for these Markovian systems, characterized by \textit{n} states and a lag time $\tau$, which records the system’s state. This matrix enables us to determine state populations and conditional pairwise transition probabilities, from which free energies are derived. We employ time-lagged independent component analysis (TICA) \cite{perez2013identification} to enhance our analysis, reducing the high-dimensional conformational space into a lower-dimensional, optimally reduced space. We then discretize this space using K-means clustering to construct the MSM. For each fast-folding protein, all-atom simulation data, characterized by pairwise C$\alpha$ distances, are projected onto the first four components using TICA. A reference free energy surface was then constructed for each system by binning the first two TICA dimensions into an 80×80 grid and averaging the weights of the equilibrium probability in each bin, as computed by the Markov state model. For the CG simulations, we adopt the approach outlined in \cite{majewski2023machine}, utilizing covariance matrices from all-atom molecular dynamics (MD) to project the first three components. This method aligns with established methodologies, ensuring consistency and facilitating further analysis. HTMD provided the necessary computational tools and framework to perform these analyses \cite{doerr2016htmd}. Finally, to avoid biasing the model with starting conformations, 10\% of the initial frames of each trajectory were removed from the analysis except for $\alpha$3D, where 5\% of the initial frames were discarded due to the longer simulation times considered.
\section{Results}
TensorNet was trained on the filtered mdCATH dataset, as described in Section \ref{sec: dataset}, using TorchMD-Net \cite{pelaez2024torchmdnet} for 100 epochs. Detailed information on model architecture and training hyper-parameters can be found in the supplementary material  (see Tables \colorit{S3} and \colorit{S4}). 

\begin{figure}
    \centering
    \includegraphics[width=\linewidth]{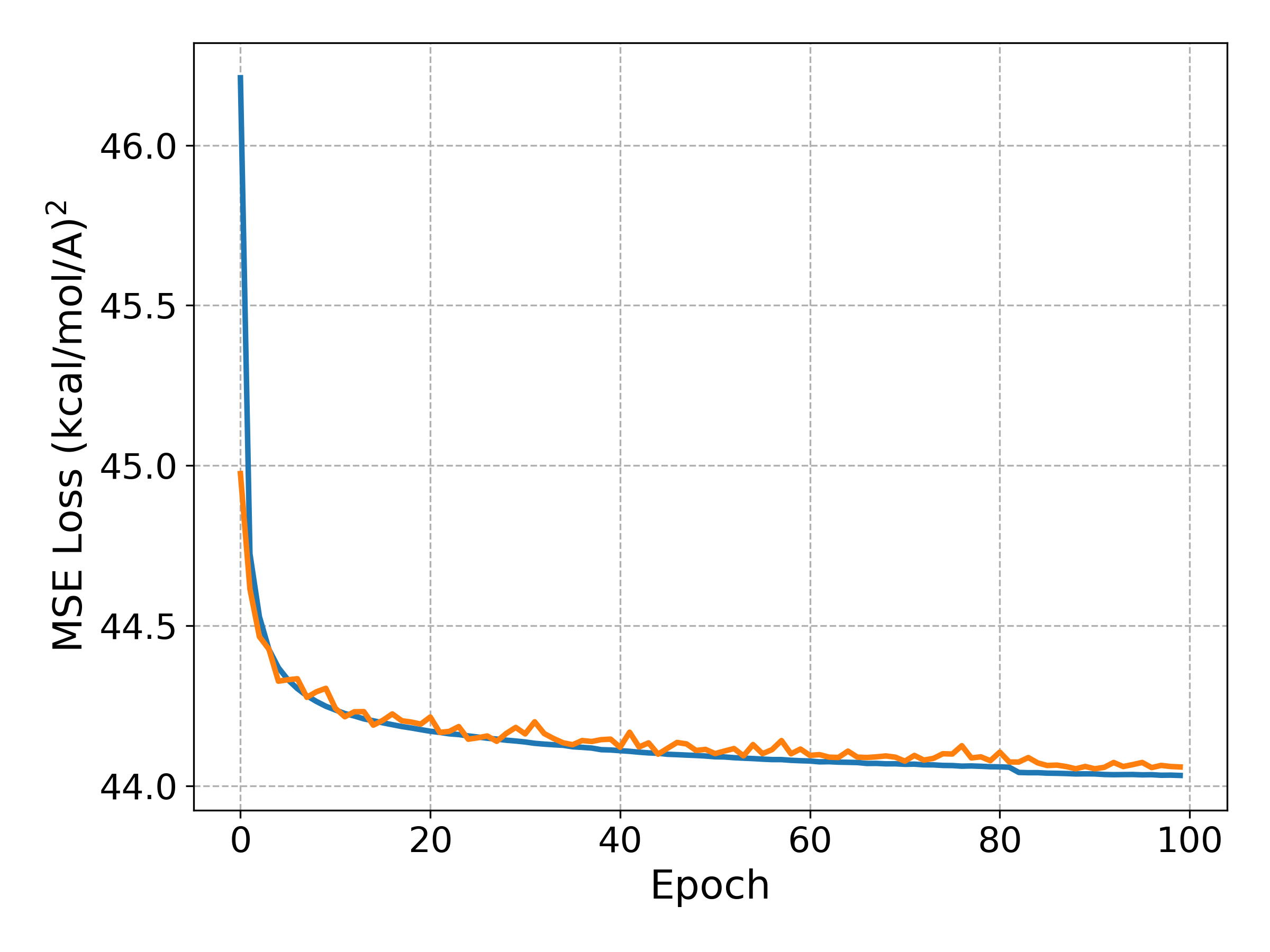}
    \caption{Traning- and validation- MSE loss, in blue and orange respectively, for \nohCGmodelName{} as a function of training epoch.}
    \label{fig: training_metrics}
\end{figure}

The final L1 test loss for the model is reported as 5.07 kcal/mol/\AA, while MSE loss for training and validation are reported in Figure \ref{fig: training_metrics}.

\subsection{Generalization to larger domains}
\begin{figure*}
    \centering
    \includegraphics[width=0.7\linewidth]{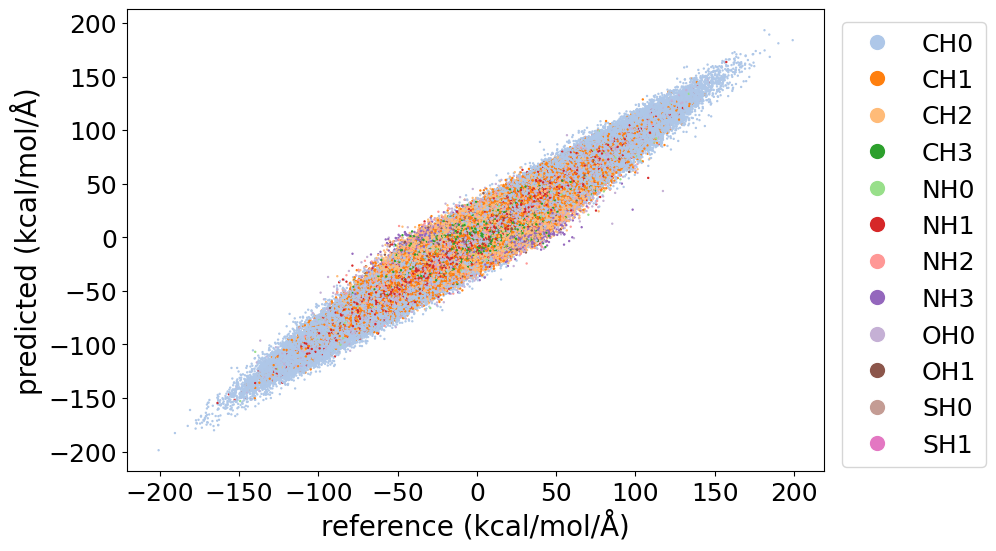}
    \caption{Scale-up validation of \nohCGmodelName{} on larger domains from mdCATH. Comparison between labeled (i.e. reference) and predicted force component values (x, y, z). Each data point in the scatter plot is color-coded according to the CG atom type.}
    \label{fig: inference_larger_domains}
\end{figure*}

We explore the ability of the \nohCGmodelName{} to scale up when larger protein domains than those considered in the training set. To achieve this, we selected a subset of 5,000 conformations from the mdCATH dataset, specifically targeting domains with between 150 and 250 residues and a combined helix and sheet fraction $>$ 50\%. This selection criteria ensured that the domains were representative of complex protein structures while still maintaining a manageable size for computational analysis. The forces acting on the noh-bead within these larger domains were evaluated using the CG-NNP. \nohCGmodelName{} exhibited a mean absolute error (MAE) of 4.98~kcal/mol/\AA, assessed for each force component (x, y, z). The error value here recorded, compared also to the one obtained at the end of the training, proves that the learned potential can scale up without loss in accuracy. Moreover, Figure \ref{fig: inference_larger_domains} presents a direct comparison between the expected and observed force values, with each dot color-coded by CG atom type, illustrating the model’s precision.
The results underscore the robustness of the CG-NNP model and its potential applicability to larger biological systems, confirming its feature to maintain performance across an expanded range of system sizes and complexities.

Detailed errors per atom type are reported in Table \colorit{S5}, with ${\text{NH}_{3}}^{+}$ exhibiting the highest error, a mean absolute error (MAE) of 8.10~kcal/mol/\AA. However, this is not due to \textit{NH3} bead-type being underrepresented in the training datasets, as shown in Figure \colorit{S1}, but rather due to the physical and chemical properties of this group. Five domains with the highest number of ${\text{NH}_{3}}^{+}$ groups (i.e. $\geq$20) were selected for further investigation: 1kvnA00, 1nu7D01, 1w9rA00, 2c5zA00, 2jzvA00, 2nc9A00 and 3qneA01. In all of these cases, the protonated amino group corresponds to the terminal amino group along the lysine side chain. These groups are oriented outward, suggesting solvent interactions, which are not accounted for in the current modeling. Moreover, in more compact structures, these groups might be involved in salt bridges with carboxyl groups. Such complex interactions contribute to the challenge of generalizing the model's parameters for this specific atom type.

\subsection{Validation on fast-folding proteins}
To assess the model's transferability, we selected four fast-folding proteins not included in the training set: Chignolin (175 atoms), Trp-Cage (210 atoms), Villin (573 atoms), and $\alpha$3D (1149 atoms); the number of atoms in their all-atom representations are indicated in parentheses. After applying the \textit{noh} mapping, the sizes were reduced to 97, 112, 286, and 
576 atoms, respectively. TorchMD \cite{doerr2020torchmd} was used to run 32 replica MD simulations for each protein and considering at least 320 ns of aggregated simulation time, see Table \colorit{S6}. The initial coordinates for these simulations were uniformly sampled from the respective TICA surfaces of each fast folder, as shown in Figure \colorit{S2}. For Trp-Cage, only residues from 2 to 16 have been considered to focus on capturing the overall folding and not the cis-trans proline (residue 20) isomerization.

\subsection{Recovering the energetic landscape}
\begin{figure*}
    \centering
    \includegraphics[width=\linewidth]{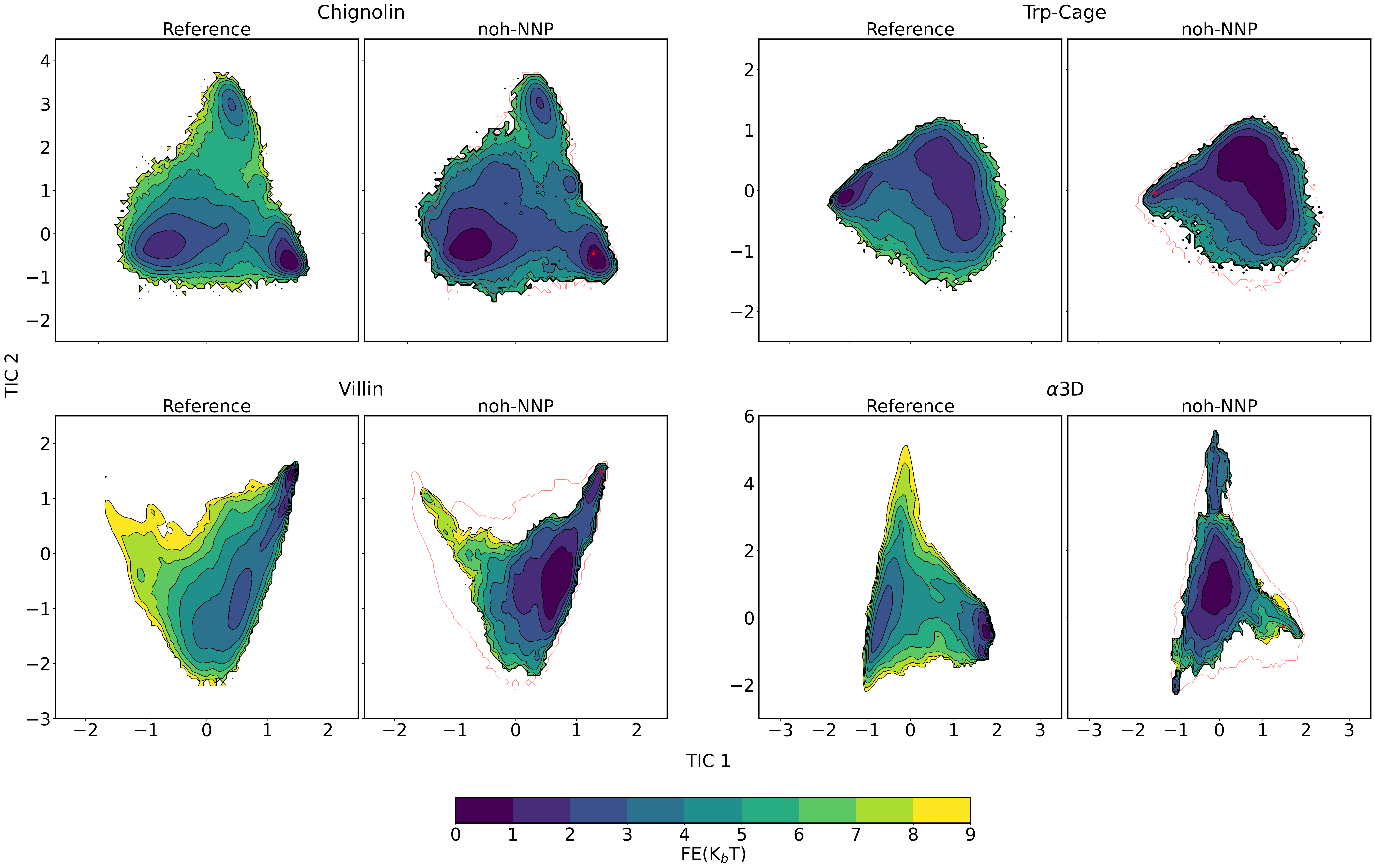}
    \caption{Comparative analysis of the free energy landscape obtained from all-atom simulations (left) and NNP coarse-grained simulations (right) across the first two TICA dimensions for four fast-folding proteins: Chignolin, Trp-cage, Villin and $\alpha$3D.}
    \label{fig: TICA}
\end{figure*}

For three of the four fast-folding proteins the TICA landscape has been successfully recovered, see Fig. \ref{fig: TICA}. In contrast, for $\alpha$3D, the recovery is only partial, with most microstates populating a middle region between the global minimum on the right and a local minimum on the left. This behavior could be attributed to the particular secondary structure of $\alpha$3D, which is characterized by a high proportion of $\alpha$-helices. Additionally, the relative shape anisotropy (RSA) of the system, 0.3, may not be well-represented in the training dataset, which has an average RSA of 0.17 ± 0.16. This discrepancy, coupled with the larger system size and the presence of ${\text{NH}_{3}}^{+}$ groups in the unstructured region between the first and second helices of the folded state, could contribute to the observed differences.

While Fig. \ref{fig: TICA} displays the two TICA dimensions as the principal axes and uses free energy as the third dimension, Fig. \colorit{S3} presents the free energy on the y-axis, providing a quantitative analysis. For the simplest target, Chignolin, the absolute minimum is perfectly captured by \nohCGmodelName{}. For more complex structures like Trp-Cage and Villin, the overall shape of the profile is well approximated. Conversely, for $\alpha$3D, extensive sampling in the central region results in a shift, leading to the identification of an absolute minimum between the two minima.

\subsection{Sampling the native structures of unseen training proteins}

\begin{figure*}
    \centering
    \includegraphics[width=\textwidth]{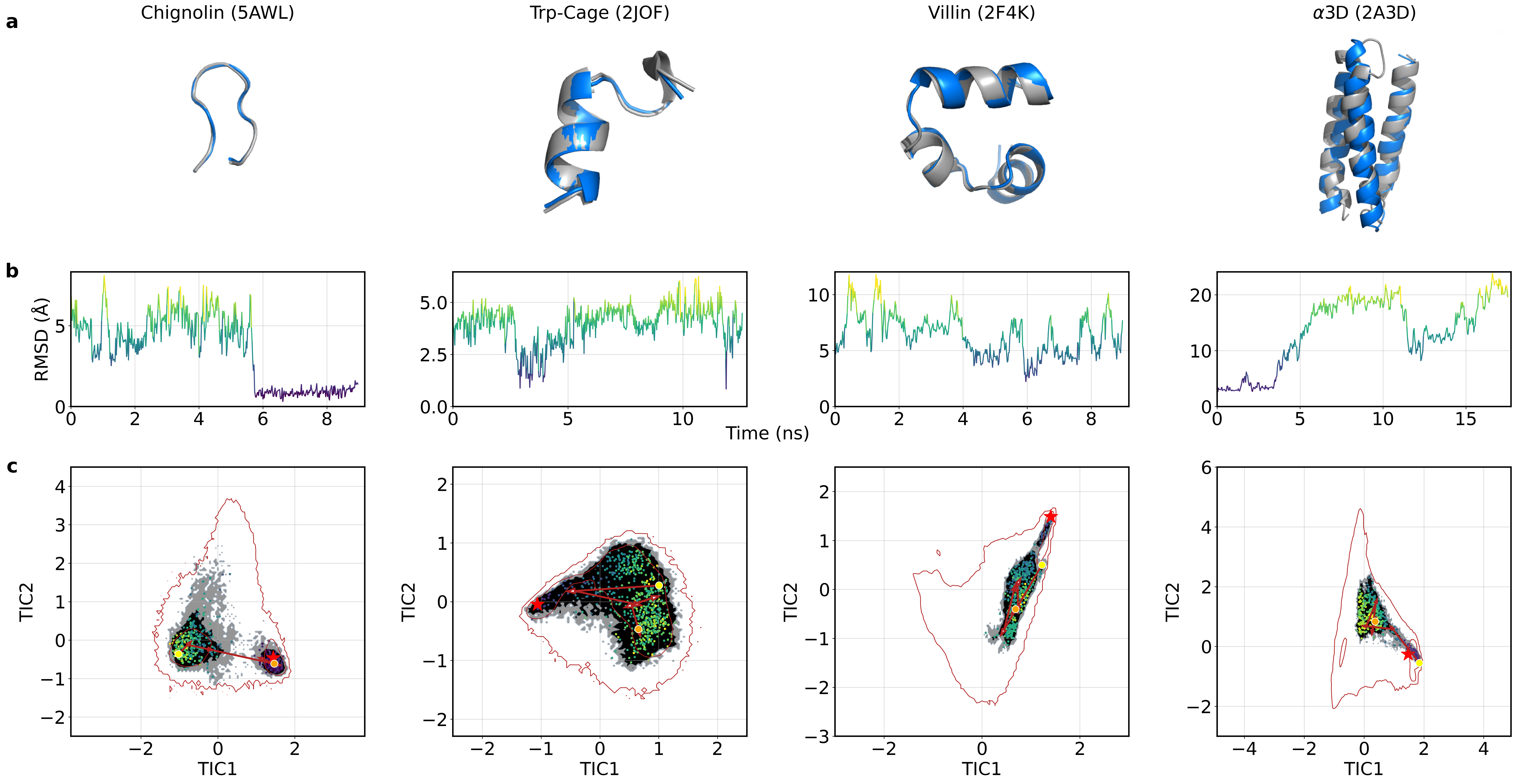}
    \caption{CG trajectories of Chignolin, Trp-Cage, Villin, and $\alpha$3D, selected based on the inclusion of microstates from the lowest RMSD macrostate. (a) Minimum RMSD conformation (blue) aligned with the experimental structure (grey) for each protein, labeled with the protein name and PDB ID. (b) C$\alpha$ RMSD of each trajectory compared to the crystal structure. (c) CG free energy surface, projected over the first two TICs with the folded state (red star) and sampled states indicated by RMSD color-coded dots. The trajectory’s progression is illustrated with arrows connecting the starting (yellow point) and ending (orange point) conformations. The all-atom equilibrium density is shown by a red contour.}
    \label{fig: native_structures}
\end{figure*}

Analysis of the CG simulations using MSMs, detailed in Section \ref{MSM_Section} and Table \colorit{S7}, revealed that the model successfully reproduced the experimental structure of the corresponding fast-folding proteins, as illustrated in Figure \ref{fig: native_structures}a. Sampling originated from the native macrostate, defined as the macrostate containing the conformation with the minimum root-mean-square deviation (RMSD) with respect to the experimental crystal structure. The models accurately predicted secondary and tertiary structural elements, with loops and unstructured terminal regions showing minimal variation, except for $\alpha$3D. Table \colorit{S8} provides detailed information on the equilibrium probability, mean, and minimum RMSDs of the native macrostates. The results indicate extensive sampling of the native conformation, as reflected by the high equilibrium probabilities for these macrostates. For Chignolin, Trp-Cage, and Villin, an average RMSD below 1 \AA\ was observed when comparing the native macrostate to the folded structure, more in detail: 0.15 \AA, 0.30 \AA, and 0.6 \AA, respectively. In contrast, the larger and more complex $\alpha$3D system reported an RMSD of 2.30 \AA. These results underscore the model's accuracy in structural prediction and its adaptability across different molecular systems, though further investigation is required to fully understand the particular case of $\alpha$3D.

\subsection{Computational efficiency}
To assess the computational efficiency of \nohCGmodelName{}, we conducted a comparative analysis against traditional all-atom simulations, focusing on sampling free energy (FE) landscapes within the TICA space. Both simulations were constrained to a 12-hour time-frame to ensure a fair comparison, with the all-atom simulations using the CHARMM22* force field \cite{mackerell1998all, mackerell2004extending}, and explicit solvent, while the \nohCGmodelName{} simulation focused on heavy atoms using a \textit{noh} mapping approach.
Both models were run on openMM \cite{eastman2023openmm} using an NVIDIA RTX 4090, and employed a Langevin thermostat at 350 K, with differing friction coefficients: 0.1 ps$^{-1}$ for CHARMM22* and 1 ps$^{-1}$ for \nohCGmodelName{} (standard setup, see Section \ref{MolSims}). Additionally, the hydrogen mass repartitioning (HMR) scheme \cite{feenstra_improving_1999} was set at 4 a.m.u. for the classical force field, and \textit{increased mass} was applied for the NNP system as detailed in the methods section. All-atom simulations employed long-range electrostatics using the particle-mesh Ewald (PME) summation method \cite{dardenParticleMeshEwald1993, harvey2009implementation} with a cutoff of 9 \AA, while van der Waals interactions used a cutoff of 9 \AA\ and a switching distance of 7.5 \AA. Hydrogen atoms were constrained using the SHAKE \cite{ryckaert1977numerical} algorithm. In contrast, \nohCGmodelName{} does not use explicit treatment for long-range interactions, as they are modeled by the NNP with a total receptive field of 10 \AA.

\begin{figure*}
    \centering
    \includegraphics[width=0.8\textwidth]{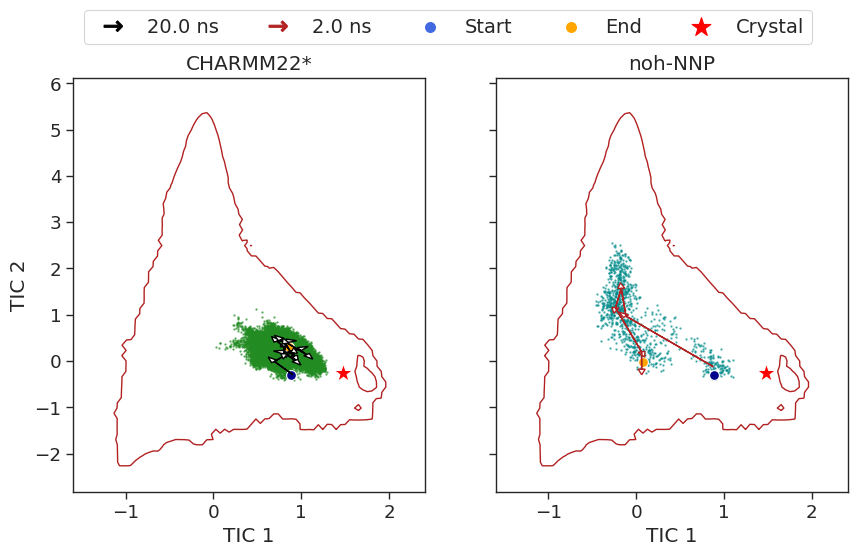}
    \caption{Sampling efficiency comparison in the TICA landscape for the $\alpha$3D system under two force fields: CHARMM22* (left) and \nohCGmodelName{} (right). Both simulations, starting from a randomly selected initial conformation (blue dot) and ending at a final conformation (orange dot), were run within a fixed 12-hour time window for direct comparison.}
    \label{fig: tica_clockTime}
\end{figure*}

The analysis revealed that the all-atom simulations achieved 265.5 ns of molecular simulation while \nohCGmodelName{} completed 8.4 ns within the same operational time window. However, to account for differences in temporal resolution between the coarse-grained and all-atom models, we compared the areas of the TICA landscapes recovered by both models, see Figure \ref{fig: tica_clockTime}. The starting point (blue dot), the same for both cases, was selected randomly and resulted in a conformation near the global minimum, while the red star represents the crystal structure (PDB ID: 2A3D). 

The all-atom simulation (left panel) explores a more localized region of the TICA space, with the trajectory (black arrows) predominantly staying within a single basin. This indicates limited sampling within the time frame due to the high dimensionality and energy barriers typical of fully atomistic representations. The final conformation is marked by a yellow dot, indicating that the simulation does not move significantly away from the initial state. 

In contrast, the CG-NNP simulation (right panel), despite the shorter trajectory length, explores a broader region of the conformational space (red arrows), as seen by the wider spread of sampled conformations. The trajectory covers a larger portion of the landscape, crossing energy barriers more easily and reaching areas of higher free energy. This behavior, characteristic of coarse-grained models, is due to the reduced degrees of freedom and the smoothing of energy landscapes which produces an effective faster kinetics. Despite this broader sampling, the path still shows consistency with the overall free energy landscape, see Figure \ref{fig: TICA} for reference, suggesting thermodynamic consistency. 

Notably, while the all-atom simulation reaches a minimum RMSD of 5.03 \AA\ with respect to the crystal structure, the CG-NNP model achieves a lower minimum RMSD of 4.38 \AA\. Moreover, the average RMSD values of 13.31 ± 4.82 \AA\ for CG-NNP and 6.41 ± 0.5 \AA\ for CHARMM22* suggest that the coarse-grained model explores a larger conformational space, as expected.

\begin{figure}
    \centering
    \includegraphics[width=\linewidth]{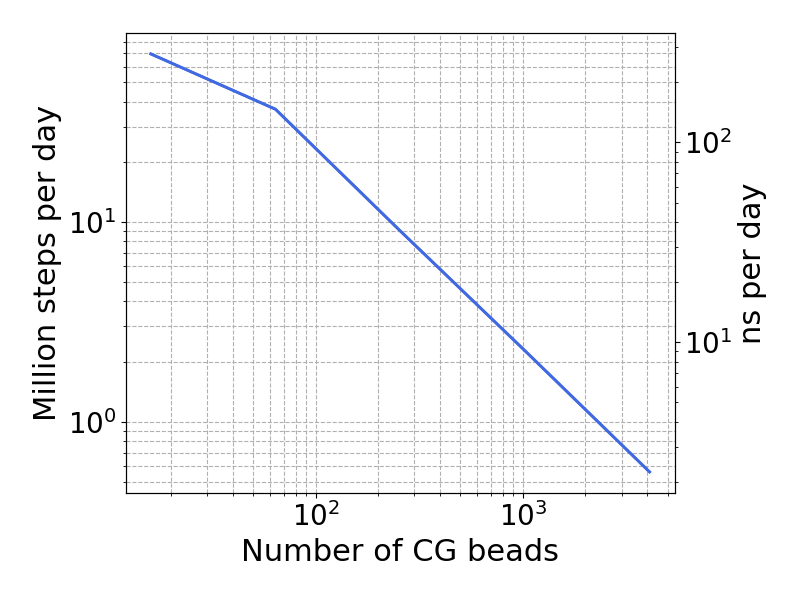}
    \caption{Log-log plot of \nohCGmodelName{}'s computational speed relative to system size (number of CG Beads). The left y-axis shows the number of simulation steps (in millions) that can be executed within a single day. The right y-axis displays the performance in nanoseconds per day (ns/day), assuming a time-step of 4 fs.}
    \label{fig:benchmark}
\end{figure}

The CG-NNP's efficiency was further quantified relative to system size, measured in CG beads (Figure \ref{fig:benchmark}, providing a quantitative benchmark of \nohCGmodelName{}'s performance. Both millions of simulation steps (left y-axis) and ns/day (right y-axis) are considered variables. The term "simulation step" refers to a forward and backward step of the model. The efficiency demonstrated by the CG-NNP in small-to-medium systems (up to ~1000 CG beads) makes it a valuable tool for exploring larger conformational landscapes or long-timescale events.

\section{Conclusions}
This paper introduces the first version of \nohCGmodelName{} a new fully machine-learning coarse-grained force field offering a new framework for molecular dynamics simulations.
\nohCGmodelName{} uses an all-heavy-atoms coarse-graining strategy paired with variational force matching, which simplifies protein representation while retaining essential dynamical information. This approach addresses previous challenges in balancing model simplicity with the retention of critical dynamics, which often require energy priors for stabilization.
Notably, our model demonstrates remarkable transferability and scaling-up ability. However, further enhancements in computational efficiency and memory usage must be achieved in the future. The current study serves as a proof of concept, challenging the reliance on prior energies in developing stable NNPs for studying protein thermodynamics.

\section{Code Availability}
Code and relevant data to reproduce this work are available at \url{https://github.com/compsciencelab/amaro}.

\section*{Associated Content}
Supporting Information is available. This section provides data and hyperparameters used in our computational experiments to ensure transparency and reproducibility. It includes specific parameters for training, model, and fast-folding protein MSMs, detailing the values used to achieve the main text results. Additional analyses and statistics are also provided.

\begin{acknowledgement}
AM is financially supported by Generalitat de Catalunya's Agency for Management of University and Research Grants (AGAUR) PhD grant 2024 FI-1-00278;  the project PID2023-151620OB-I00 has been funded by MCIN / AEI / 10.13039/501100011033. Research reported in this publication was partially supported by the National Institute of General Medical Sciences (NIGMS) of the National Institutes of Health under award number R01GM140090. 
The content is solely the responsibility of the authors and does not necessarily represent the official views of the National Institutes of Health.
\end{acknowledgement}
\clearpage
\providecommand{\latin}[1]{#1}
\makeatletter
\providecommand{\doi}
  {\begingroup\let\do\@makeother\dospecials
  \catcode`\{=1 \catcode`\}=2 \doi@aux}
\providecommand{\doi@aux}[1]{\endgroup\texttt{#1}}
\makeatother
\providecommand*\mcitethebibliography{\thebibliography}
\csname @ifundefined\endcsname{endmcitethebibliography}  {\let\endmcitethebibliography\endthebibliography}{}

\end{document}


\date{}
\maketitle
\clearpage
\begin{table}
    \centering
    \begin{tabular}{|c|c|c|c|}
    \hline
    \textbf{Group Type} & \textbf{Embedding Value}\\
    \hline
     C &  1 \\
     CH & 2 \\
     $\text{CH}_{2}$ & 3 \\
     $\text{CH}_{3}$ & 4 \\
     N & 5 \\
     NH & 6 \\
     $\text{NH}_{2}$ & 7 \\
     $\text{NH}_{3}$ & 8 \\
     O & 9 \\
     OH & 10 \\
     S & 11 \\
     SH & 12 \\
     \hline
\end{tabular}
\caption{Embedding values for \textit{noh}-beads, based on the heavy atom and the number of bonded hydrogen atoms.}
\label{tab: embeddings}
\end{table}

\begin{table}
    \centering
    \begin{tabular}{|c|c|c|}
    \hline
        Target & Target Sequence Length & Sequence Overlap (\%) \\
        \hline
        Chignolin (5AWL) & 10 & 60  \\
        Trpcage (2JOF) & 20 & 55 \\
        Villin (2F4K) & 35 & 51 \\
        $\alpha$3D (2A3D)& 73 & 33 \\ 
        \hline
    \end{tabular}
    \caption{Maximum sequence similarity, $\sigma$, for fast-folding proteins compared to the mdCATH train/val/test dataset. Given a target sequence (ST) and a reference sequence (SR), then $\sigma =$(Num. Matching Residues)/$|\text{Align(ST, SR)}|$, where "Align" is a lexicographic alignment function and $|\cdot|$ represents the length of the alignment. Alignments were performed using Biopython's pairwise2.align.localxs function \cite{cock2009biopython}, with both gap open and extension penalties set to -1.}
    \label{tab: sequence_similarity}
\end{table}

\begin{table}
    \centering
    \begin{tabular}{|c|c|}
    \hline
    \textbf{Hyper-parameter} & \textbf{Value} \\
    \hline
    activation & Silu \\
    aggr & add \\
    cutoff\_lower & 0.0 \AA \\
    cutoff\_upper & 5.0 \AA \\
    embedding\_dimension & 128 \\
    equivariance\_invariance\_group & O(3) \\
    max\_num\_neighbors &  64 \\
    max\_z & 100 \\
    num\_layers & 1.0 \\
    num\_rbf & 32 \\
    precision & float32 \\
    rbf\_type & expnorm \\
    \hline
\end{tabular}
\caption{Neural network architecture hyperparameters}
\label{tab: tensornet_hyprmts}
\end{table}

\begin{table}
    \centering
    \begin{tabular}{|c|c|}
    \hline
    \textbf{Hyper-parameter} & \textbf{Value} \\
    \hline
    optimizer & AdamW  \\
    batch size & 8 \\
    distance\_influence & both \\
    early\_stopping\_patience & 30 \\
    ema\_alpha\_neg\_dy & 1.0\\ 
    ema\_alpha\_y & 1.0 \\ 
    learning rate & 0.0003 \\
    max\_num\_epochs & 100 \\
    neg\_dy\_weight & 1.0 \\
    test\_interval & -1 \\
    test\_size & 0.1 \\
    train\_size & null \\
    trainable\_rbf & false \\
    val\_size & 0.05 \\
    weight\_decay &  null \\
    y\_weight & 0.0 \\
    \hline
\end{tabular}
\caption{Neural network training hyperparameters}
\label{tab: train_tensornet_hyprmts}
\end{table}

\begin{table}
    \centering
    \begin{tabular}{|c|c|}
        \hline
         CG-atom type & MAE (kcal/mol/\AA)  \\
         \hline
         CH0 & 5.56 \\
         CH1 & 5.02 \\
         CH2 & 6.05 \\ 
         CH3 & 5.44 \\ 
         NH0 & 5.17 \\ 
         NH1 & 4.27 \\ 
         NH2 & 5.30 \\ 
         NH3 & 8.10 \\ 
         OH0 & 3.53 \\ 
         OH1 & 5.09 \\ 
         SH0 & 4.79 \\ 
         SH1 & 3.54 \\
         \hline
    \end{tabular}
    \caption{Mean Absolute Error (MAE) for the x, y, and z components of forces predicted by \nohCGmodelName{}. The errors are classified by atom type across a test set of 5,000 conformations, each containing more than 150 residues, and a range of 5 temperatures. The overall MAE across all components and atom types is 4.98 kcal/mol/\AA.}
    \label{tab: at_scale_up_errors}
\end{table}

\begin{figure}
    \centering
    \includegraphics[width=0.8\linewidth]{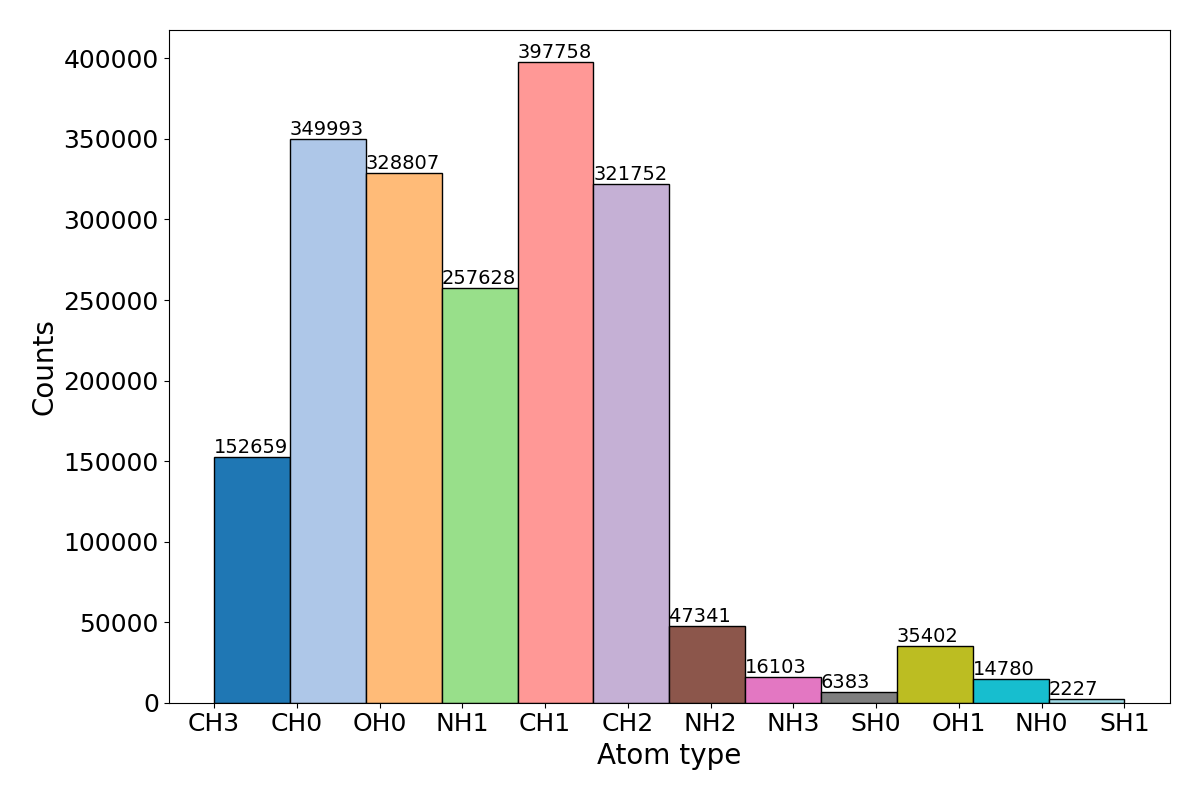}
    \caption{Distribution of coarse-grained atom types across domains in the training dataset.}
    \label{fig: dataset_atomtype_distribution}
\end{figure}

\begin{figure*}
    \centering
    \includegraphics[width=\textwidth]{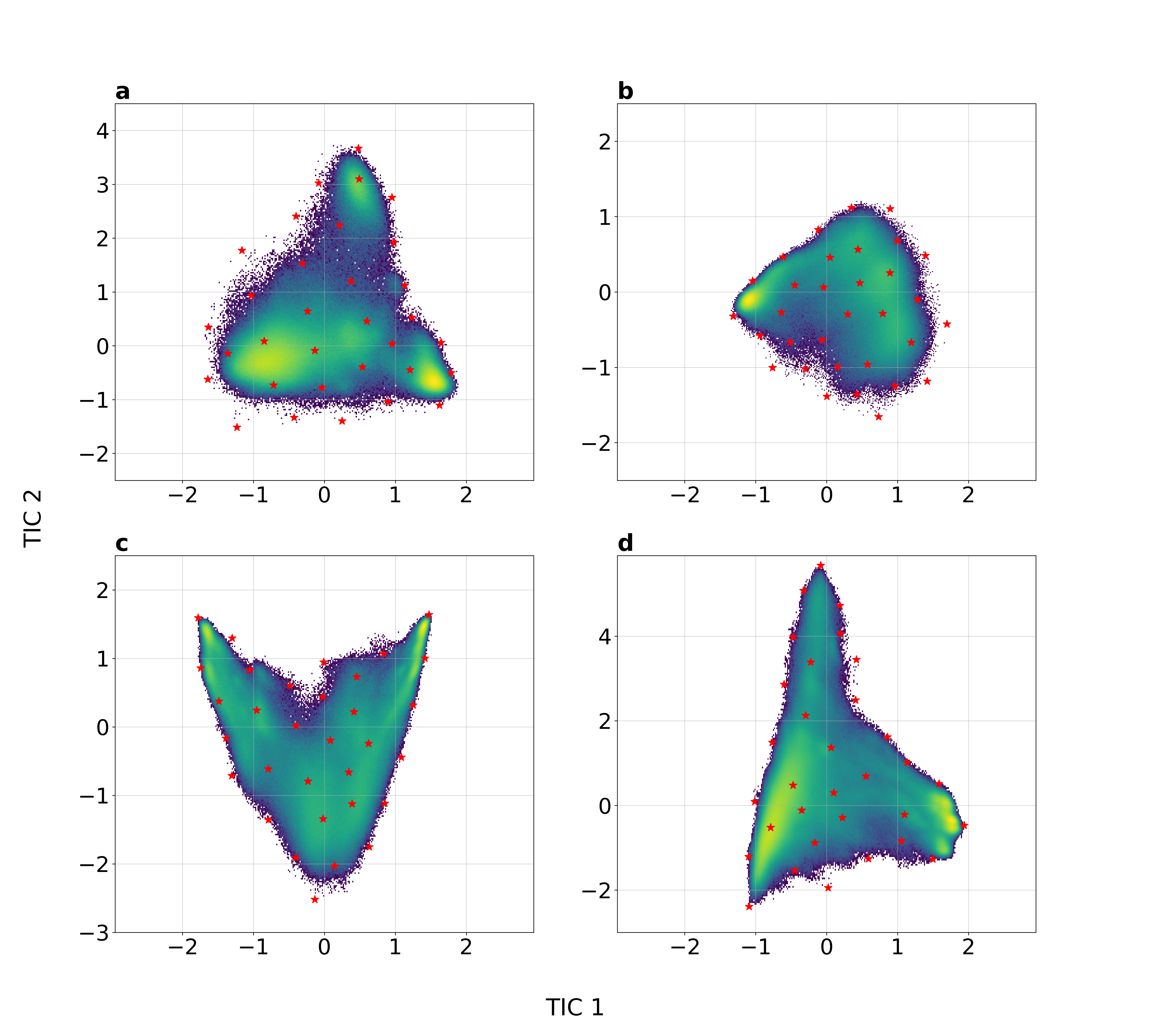}
    \caption{Initial configurations of fast-folding proteins in the test set, used to launch 32 uniformly distributed coarse-grained replica simulations across the TIC1-TIC2 space. More in details, a) Chignolin, b) Trp-cage, c) Villin, and d) $\alpha$3D.}
    \label{fig: start_coords_xtc}
\end{figure*}

\begin{table}
\centering
\begin{tabular}{|c|c|c|c|}
\hline
\textbf{protein} & \textbf{Numb. of steps} & \textbf{Median length (ns)} & \textbf{Aggr. length (ns)}  \\
\hline
Chignolin & 2.5M & 10 & 320 \\
\hline
Trp-Cage & 3.5M & 14  & 448  \\
\hline
Villin & 2.5M & 10 & 320\\
\hline
$\alpha$3D & 4.5M & 18 & 576 \\
\hline
\end{tabular}
\caption{Summary of the lengths of 32 MD trajectory replicas for each fast-folding protein used in the MSM analysis.}
\label{tab: ff_sims}
\end{table}

\begin{table}
\centering
\begin{tabular}{|c|c|c|c|}
\hline
\textbf{protein} & \textbf{numCluster} & \textbf{numMacordim} & \textbf{lag time (ns)} \\
\hline
Chignolin & 350 & 3 & 0.1  \\
\hline
Trp-Cage & 350 & 2 & 0.5 \\
\hline
Villin & 250 & 3 & 0.5 \\
\hline
$\alpha$3D & 600 & 3 & 0.5\\
\hline
\end{tabular}
\caption{Parameters used to construct MSMs for analyzing the dynamics of different fast-folding protein trajectories simulated under \nohCGmodelName{}.}
\label{tab: MSM_hparms}
\end{table}

\begin{table}
    \centering
    \resizebox{\textwidth}{!}{
    \begin{tabular}{|c|c|c|c|c|c|c|}
        \hline
        Protein & \multicolumn{3}{c|}{CG-NNP} & \multicolumn{3}{c|}{all-atom} \\
        \cline{2-7}
        & Min RMSD (\AA) & Mean RMSD (\AA) & Macro Prob. (\%) & Min RMSD (\AA) & Mean RMSD (\AA) & Macro Prob. (\%) \\
        \hline
        Chignolin & 0.16 & 1.11 $\pm$ 0.5 & 26.93 & 0.15 & 1.02 $\pm$ 0.4 &  57.53\\
        \hline
        Trp-Cage & 0.31 & 3.61 $\pm$ 0.8 & 63.6 & 0.45 & 2.46 $\pm$ 0.82& 30.1 \\
        \hline
        Villin & 0.6 & 2.7 $\pm$ 0.9 & 4.5 & 0.47 & 3.44 $\pm$ 1.84 & 69.42 \\
        \hline
        $\alpha$3D & 2.30 & 3.30 $\pm$ 0.51 & 1.2 & 1.81 & 3.50 $\pm$ 0.75 & 67.89\\
        \hline
    \end{tabular}
    }
    \caption{Minimum Average RMSD Macrostate Statistics derived from Markov State Models (MSM) built with coarse-grained simulations and all-atom molecular dynamics for fast-folding proteins. The table displays the average and minimum RMSD values (in \AA) for each macrostate alongside its equilibrium probabilities, expressed as percentages (macro prob.).}
    \label{tab: msm_statistics}
\end{table}

\begin{figure*}
    \centering
    \includegraphics[width=\textwidth]{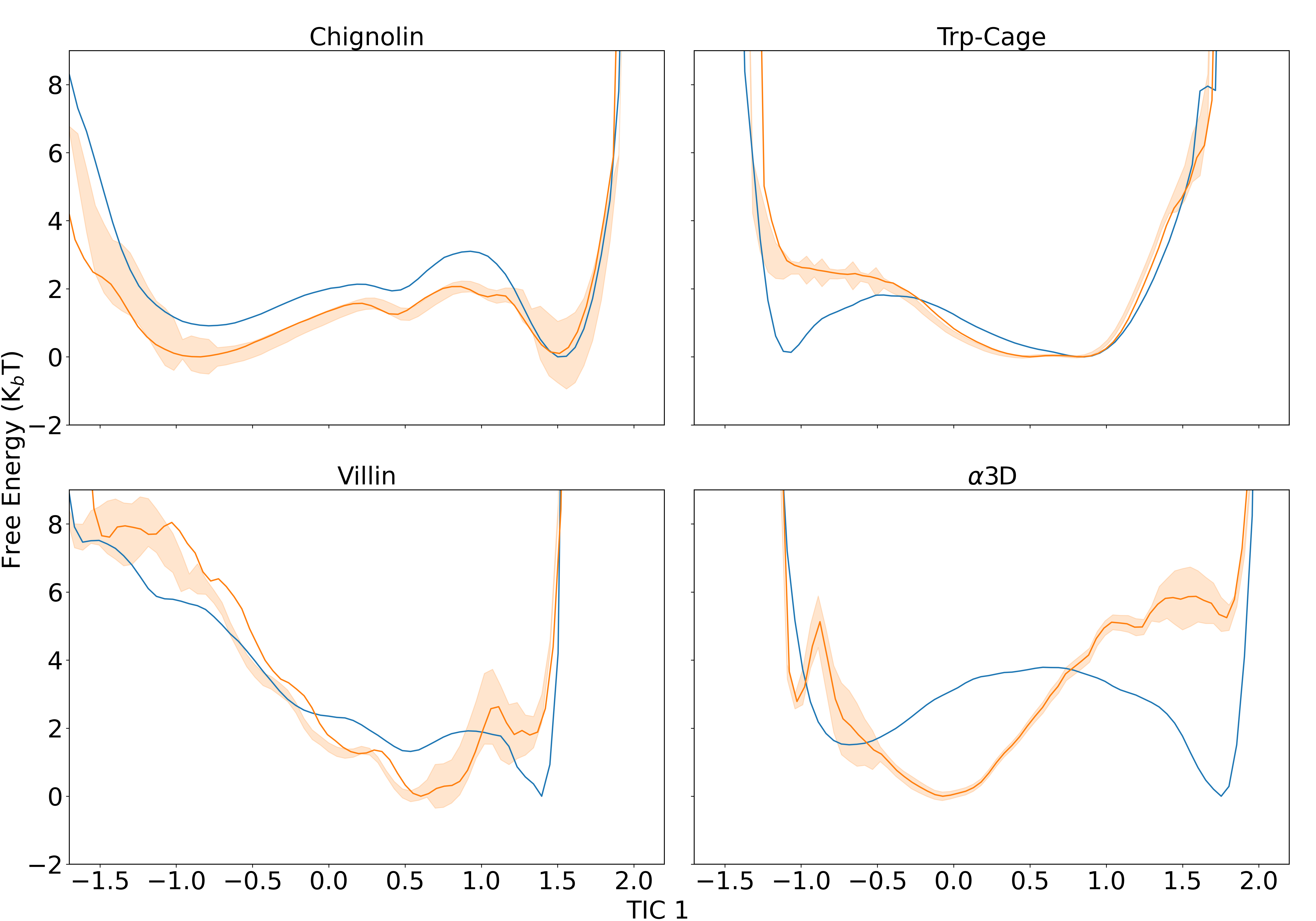}
    \caption{Free Energy comparison along the slowest TIC between NNP coarse-grained simulations (in orange) and the relative all-atom MD simulation (in blue). The standard deviation weighted over the number of replicas for \nohCGmodelName{} is reported as shaded filling.}
    \label{fig: FE_TIC1}
\end{figure*}

\clearpage
\bibliographystyle{achemso}
\providecommand{\latin}[1]{#1}
\makeatletter
\providecommand{\doi}
  {\begingroup\let\do\@makeother\dospecials
  \catcode`\{=1 \catcode`\}=2 \doi@aux}
\providecommand{\doi@aux}[1]{\endgroup\texttt{#1}}
\makeatother
\providecommand*\mcitethebibliography{\thebibliography}
\csname @ifundefined\endcsname{endmcitethebibliography}  {\let\endmcitethebibliography\endthebibliography}{}